# Classification of graphs based on homotopy equivalence. Homotopy equivalent graphs. Basic graphs and complexity of homotopy equivalence classes of graphs.

Alexander V. Evako

OOO "Dianet", Laboratory of Digital Technologies.

Volokolamskoe Sh. 1, kv. 157, 125080, Moscow, Russia

Tel/Fax: +7 499 158 2939, e-mail: evakoa@mail.ru.

**Abstract**

Graph classification plays an important role is data mining, and various methods have been developed recently for classifying graphs. In this paper, we propose a novel method for graph classification that is based on homotopy equivalence of graphs. Graphs are called homotopy equivalent if one of them can be converted to the other one by contractible transformations. A basic graph and the complexity of a homotopy equivalence class are defined and investigated. It is shown all graphs belonging to a given homotopy equivalence class have similar topological properties and are represented by a basic graph with the minimal number of points and edges. Diagrams are given of basic graphs with the complexity $N \leq 6$.

The advantage of this classification is that it relies on computer experiments demonstrating a close connection between homotopy equivalent topological spaces and homotopy equivalent graphs.

# 1. Introduction

Graphs are widely used to represent the structure of discrete images arising in many areas of science including chemistry, biology, medicine, computer graphics and so on. In neuroscience, generating topologically correct discrete models of anatomical structures is critical for many clinical and research applications [1]. Over the past few decades, there has been a growth of interest in studying structural properties of graphs and various classifications which separate graphs into classes. Many interesting results have been received in this area by researchers (recall, for example, [2, 3, 4], etc.). Paper [5] proposes an efficient method for mining discriminative subgraphs for graph classification in large databases. A novel approach to graph classification that is based on feature vectors constructed from different global topological attributes, as well as global label features was investigated in [6]. Homotopy features of graphs have been studied in a number of works. (see e.g.[7-11]). One of the goals for this study is to find out structural invariants for dynamic processes that can be encoded into graphs. It was shown in [12, 13], that transformations of graphs called contractible preserve topological features of graphs such as the Euler characteristic and the homology groups of graphs. This result leads us to understanding that contractible transformations define homotopy equivalence between graphs.

The classification of graphs proposed in the present paper is different from all types of classifications considered before. The major contribution of this paper is that it provides an efficient method to classify graphs with respect to their topological properties in such a way that graphs with similar topological attributes belong to the same homotopy equivalence class. The advantage of this classification is that it is analogous to classification of topological spaces by a homotopy type in algebraic topology and relies on computer experiments demonstrating a close connection between homotopy equivalent topological spaces and homotopy equivalent graphs. All graphs belonging to a given homotopy equivalence class hold identical topological attributes, and are represented by a basic graph with the minimal number of points and edges. Instead of studying topology of a large graph it is much easier to study topology of its basic graph.

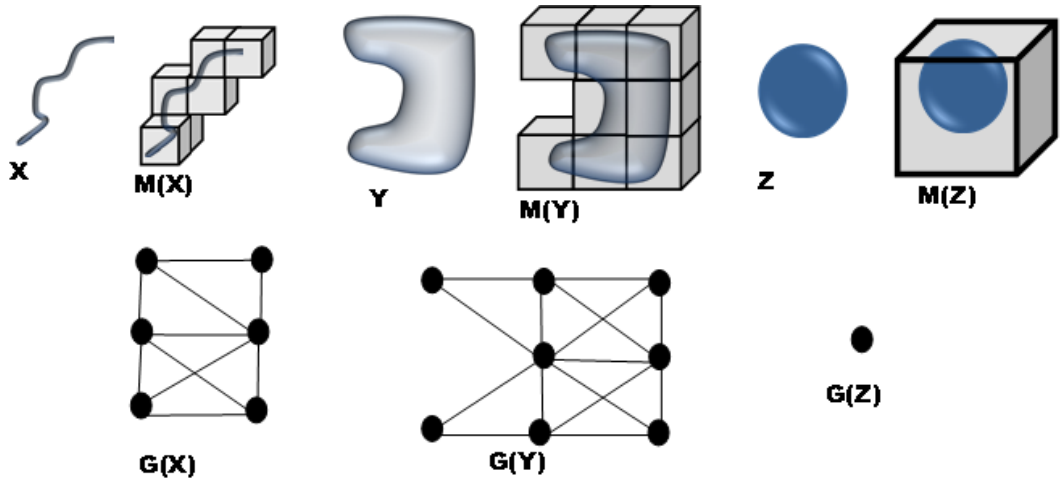

Figure 1 X is a twisted curve, M(X) is the set of cubes intersecting X, G(X) is the intersection graph of M(X). Y is a topological disk, M(Y) is the set of cubes intersecting Y, G(Y) is the intersection graph of M(Y). Z is a topological ball, M(Z) is the set of cubes intersecting Z, G(Z) is the intersection graph of M(Z). Spaces X, Y and Z are homotopy equivalent, and graphs G(X), G(Y) can be converted with contractible transformations to each other and to a one-point graph G(Z).

This paper is organized as follows. Computer experiments are presented in section 2. In section 3, we describe basic definitions and results related to the topic of this paper and obtained in previous works. New results are presented in sections 4-6. We study new properties of contractible transformations and homotopy equivalent graphs. We give the definition of a simple closed curve and a simply connected graph and study properties of simply connected graphs. We define a simple subgraph of a given graph and show that a simple subgraph can be contracted to a point without changing the homotopy type of the graph. The definitions of a compressed minimal graph and a basic graph of a homotopy equivalence class are given, and it is shown that any graph belonging to a given homotopy equivalence class can be converted to the basic graph by sequential deleting simple points, simple edges and by the sequential contraction of simple sets of points. Since all homotopy equivalent graphs are characterized by a similar topological structure, it is enough to investigate structural properties of only basic graphs. All graphs are classified by the basic graphs, the complexity and the weight. Basic graphs of homotopy equivalence classes with complexity ≤ 6 are presented.

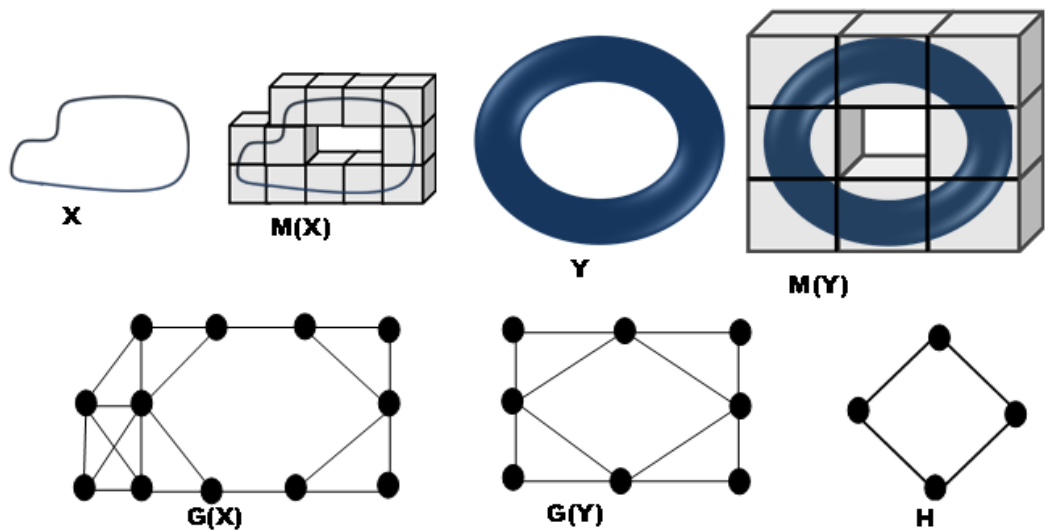

Figure 2   X is a circle, M(X) is the set of cubes intersecting X, G(X) is the intersection graph of M(X). Y is a solid torus, M(Y) is the set of cubes intersecting Y, G(Y) is the intersection graph of M(Y). Spaces X and Y are homotopy equivalent, and graphs G(X), G(Y) can be converted with contractible transformations to each other and to graph H, which is a minimal digital 1-sphere.

## 2 Computer experiments

The following interesting facts were observed in computer experiments described in [12]. Let space X be located in Euclidean space $E^n$. Divide $E^n$ into the set of n-cubes with the side length L and vertex coordinates in the set X={ $Lx_1,…Lx_n$: $x_i \in Z$, for i=1,...,n}. Call the cubical model of X the family M(L,X) of all n-cubes intersecting X, and the digital model of X the intersection graph G(L,X) of M(L,X).

Suppose that spaces X and Y are of the same homotopy type. It was revealed that $L_0>0$ exists such that for all positive $L_1<L_0$ and $L_2<L_0$, graphs $G(L_1,X)$ and $G(L_2,Y)$ can be transformed from one to the other by a special type of transformations, called contractible. For example, if X and Y are contractible spaces then

$G(L_1,X)$ can be converted to $G(L_2,Y)$ by contractible transformations, and both $G(L_1,X)$ and $G(L_2,Y)$ can be converted to a trivial one-point graph by contractible transformations.

These results of experiments show that:
- If continuous spaces are homotopy equivalent then their intersection graphs are homotopy equivalent.
- The intersection graph G(X) contains topological and perhaps geometrical characteristics of continuous space X. Otherwise, the digital model G(X) is a digital counterpart of X.
- Contractible transformations of graphs are a digital analogue of homotopy in algebraic topology.

To illustrate these experiments, consider examples depicted in fig. 1 and 2. For a twisted curve X shown in fig. 1, M(X) is a set of cubes intersecting X, and G(X) is the intersection graph of M(X). For a topological disk Y, M(Y) is a set of cubes intersecting Y, and G(Y) is the intersection graph of M(Y). For a topological ball Z, M(Z) is a set of cubes intersecting Z, and G(Z) is the intersection graph of M(Z). Obviously, spaces X, Y and Z are homotopy equivalent, and graphs G(X), G(Y) and G(Z) can be transformed with contractible transformations to each other and to a one-point graph G(Z). In fig. 2, X is a circle, M(X) is a set of cubes intersecting X, and G(X) is the intersection graph of M(X). Y is a solid torus, M(Y) is a set of cubes intersecting Y, and G(Y) is the intersection graph of M(Y). Spaces X and Y are homotopy equivalent. It is easy to check that G(X) can be converted to G(Y) by contractible transformations. Similarly, G(X) and G(Y) can be converted to a graph H by contractible transformations. H is a minimal digital 1-dimensional sphere according to [12]. Properties of graphs representing digital n-manifolds and surfaces were investigated in [14, 15].

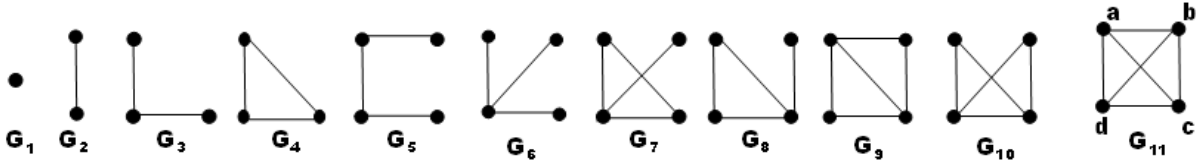

Figure 3. Contractible graphs with the number of points n<5.

## 3 Preliminaries

In order to make this paper self-contained we will summarize the necessary information from papers [12, 13, 14, 16]. By a graph G=(V,W), we mean a finite or countable set of points V={$v_1,v_2,...v_n,...$} together with a set of edges W = {$(v_pv_q)$,....}⊆V×V, provided that $(v_pv_q)=(v_qv_p)$ and $(v_pv_p)\notin W$. Remind that a subgraph H=($V_1,W_1$) of a given graph G=(V,W) is said to be induced by the set of points $V_1$ if points $v_p,v_q\in H$ are adjacent in H if and only if they are adjacent in G. Since in this paper we use only subgraphs induced by a set of points, we use the word subgraph for an induced subgraph. We write H⊆G. In [12], the subgraph O(v)⊆G containing all points adjacent to v (without v) is called the rim or the nearest neighborhood of point v in G. The subgraph B(v)=v∪O(v) containing v as well as O(v) is called the ball of point v in G. Let (vu) be an edge of G. The subgraph O(vu)=O(v)∩O(u) is called the rim of (vu). Graphs G and H are called separated if they have no points in common and any point of G is not adjacent to any point of H. For two separated graphs, their join G⊕H is the graph that contains G, H and edges joining every point in G with every point in H. A graph v⊕G is called a cone graph. Graphs can be transformed from one into another in a variety of ways. Contractible transformations of graphs play the same role in this approach as a homotopy in algebraic topology. First, define contractible graphs.

**Definition 3.1**
- A one-point graph is contractible. If G is a contractible graph and H is a contractible subgraph of G then G can be converted into H by sequential deleting simple points.
- A point v in graph G is simple if the rim O(v) of v is a contractible graph.
- An edge (uv) of a graph G is called simple if the rim O(vu)=O(v)∩O(u) of (uv) is a contractible graph.

By construction, a contractible graph is connected. Contractible graphs with the number of points n<5 are shown in fig. 3. A contractible graph can be converted to a point by sequential deleting simple points. For example, $G_{11}$ is converted to $G_1$ by deleting points d, b and c. Every point of graphs $G_9 \div G_{11}$ in fig. 3 is simple. The following properties of graphs that we will need in this paper were studied in [12, 13, 14].

**Theorem 3.1**
(a) Let G be a contractible graph and v be a point in G. Then G can be transformed into v by sequential deleting simple points.
(b) Let G be a contractible graph and (uv) be a simple edge in G. Then G-(uv) obtained by

deleting (uv) from G is a contractible graph.
(c) Let G be a contractible graph of order |G|>1. Then G contains at least two simple points.
(d) A cone graph v⊕G is contractible.

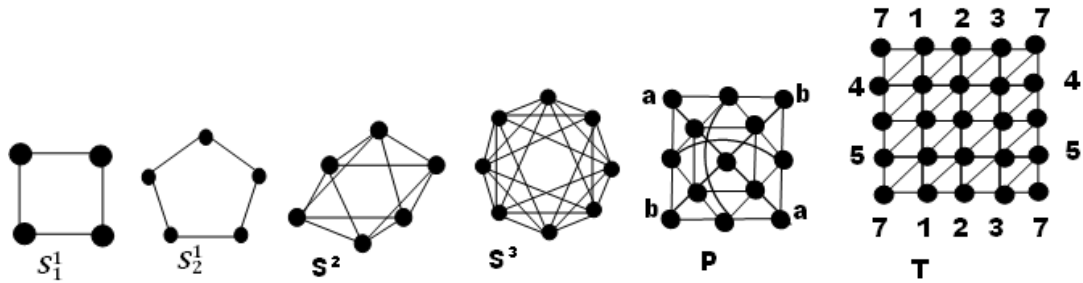

Figure 4. Graphs $S_1^1$, $S^2$, $S^3$, P, T, are compressed. $S_1^1$ and $S_2^1$ are simple closed curves. Graphs $S_1^1$, $S^2$ and $S^3$ are minimal digital 1-, 2- and 3-spheres, P is a minimal digital 2-dimensional projective plane and T is a minimal digital 2-dimensional torus.

(e) Let G be contractible graphs and H be its contractible subgraph. If all simple points of G belong to H then H=G.
(f) Let G be a contractible graph with the set of points $\{v_1,…v_n\}$ and K(n) be an n-clique with the same set of points. Then K(n) can be converted into G by sequential deleting simple edges.

Assertion (f) in theorem 3.1 is illustrated in fig. 3. $G_{11}$=K(4) is a clique. $G_{10}=G_{11}$-(ab), $G_9=G_{11}$-(bd), $G_8=(G_{11}$-(ab))-(bd), $G_7=(G_{11}$-(ab))-(bc), $G_6=((G_{11}$-(bc))-(ab))-(ac), $G_5=((G_{11}$-(bc))-(ac))-(bd).

**Definition 3.2**
- Deletions and attachments of simple points and edges are called contractible transformations.
- Graphs G and H are called homotopy equivalent if one of them can be converted to the other one by a sequence of contractible transformations.
- A graph G is called minimal if the order and the size of G cannot be reduced by contractible transformations.

Homotopy equivalent graphs have the same homotopy type and belong to the same homotopy class. All contractible graphs are homotopy equivalent to a one-point graph (see fig. 3). In papers [12, 13], it was shown that graphs, that are homotopy equivalent, have isomorphic homology groups and the Euler characteristic is also a homotopy invariant.

**4 Simply connected graphs, simple subgraphs, and compressed graphs**

In algebraic topology, the notion of a simple closed curve is used in the Poincaré conjecture. In the framework of digital topology, it was shown in [17] that a digital simple closed curve of more than 4 points is not contractible in $Z^n$. We define a simple closed curve on a graph as follows.

**Definition 4.1**
We say that a simple closed curve L is an alternating sequence $\{v_1,v_2,...v_n\}$ of different points with each point being adjacent exactly to two nonadjacent points immediately preceding and succeeding it in the sequence and with endpoints $v_1$ and $v_n$ being adjacent.

Notice that the number of points n>3. In [15, 18], a simple closed curve is called a digital one-dimensional sphere. Graphs $S_1^1$ and $S_2^1$ in fig. 4 are simple closed curves.

**Definition 4.2**
A graph G is simply connected if it is connected, and whenever L is a simple closed curve in G, then there is a contractible subgraph of G containing L.

The following theorem describes the structure of a simply connected graph.

**Theorem 4.1**
Let G be a simply connected graph. Then for any simple closed curve L lying in G there is a contractible subgraph H of G such that L is a subgraph of H, every point v belonging to L is simple in H and every point u belonging to H-L is not simple in H.

Proof.

According to definition 4.2, G contains a contractible subgraph A, which contains L. Sequentially delete from A simple points, which do not belong to L. In the obtained contractible graph H, any point u belonging to H-L is not simple. Consider a point v belonging to L. By construction of L, L-v is a contractible graph. Therefore, H can be transformed into L-v by sequential deleting simple points according to definition 2.1. Since any point belonging to H-L is not simple then v must be a simple point in H. This completes the proof. □

Graphs $S^2$ and $S^3$ depicted in fig. 4 are digital 2- and 3-dimensional spheres studied in [15, 18]. It is easy to see that they are simply connected. Graph P is a digital 2-dimensional projective plane, graph T is digital 2-dimensional torus. P and T are not simply connected. Consider properties of contractible graphs that we will use below.

**Theorem 4.2**

The join G⊕H of a contractible graph G and a graph H is a contractible graph.

Proof.

The proof is by induction on order |G|. For |G|=1,2 the theorem is plainly true. Assume that the theorem is valid whenever |G|<k. Let |G|=k. Since G is contractible then it has a simple point x. Then the rim O(x) of x in G⊕H is O(x)=O(x)$_G$⊕H, where O(x)$_G$ is the rim of x in G. Since O(x)$_G$ is contractible then O(x) is contractible by the induction hypothesis. Therefore, x is a simple point in G⊕H and can be deleted. Graph (G-x)⊕H is contractible by the induction hypothesis. Hence, G⊕H is a contractible graph by definition 2.1. This completes the proof. □

**Theorem 4.3**

Let G be a contractible graph, x∈G and B(x) is the ball of x. Then subgraph H=G-B(x) contains at least one simple point.

Proof.

The ball U(v)=v⊕O(v) is a contractible graph by proposition 2.2. Therefore, G can be converted into U(v) by sequential deleting simple points a, b,... according to corollary 2.1. Hence, 'a' is a simple point belonging to H. This completes the proof. □

**Remark 4.1**

It follows directly from this theorem that a contractible graph G is homotopy equivalent to the ball B(v) of any point v belonging to G.

Computer experiments described in [12] show that graphs corresponding to homotopy equivalent continuous spaces are homotopy equivalent to one another. This is a good argument to believe that topological and, perhaps, geometrical characteristics of continuous spaces can be encoded into graphs. Since contractible graphs play a fundamental role in this study it is desirable to construct graphs which are counterparts of topological n-dimensional disks. In literature, a graph without simple points and simple edges is called a skeleton. In paper [14], it was shown that any graph can be transformed to a skeleton by two steps: firstly we sequentially delete all simple edges, then we delete all simple points which are hanging ones. At the same time, the number of points and edges in a skeleton also can be reduced. The basic idea of a simple point can be extended to the notion of a simple subgraph. Using analogy with paper [18], we give the following definition of a simple subgraph.

**Definition 4.3**

Let G be a graph and S with points {$v_1,v_2,…v_n$} be a subgraph of G.
- S is said to be simple if the union B(S)=B($v_1$)∪…B($v_n$) of balls of all points in S is a contractible graph.
- The contraction of a simple graph S to a point z is the replacement of points {$v_1,v_2,…v_n$} with a point z such that z is adjacent to the points to which points {$v_1,v_2,…v_n$} were adjacent. The graph obtained by this contraction is denoted by H=(G-S)∪{z}.

Obviously, the contraction of a simple subgraph reduces the order of a graph.

**Theorem 4.4**

Let G be a graph and S={$v_1,v_2,…v_n$} be a simple subgraph of G. Then graph H=(G-S)∪{z} obtained by contracting S to z is homotopy equivalent to G.

Proof.

Since B(S)=U($v_1$)∪…U($v_n$) is a contractible graph, we can glue a simple point z to G in such a way that O(z)=B(S). The obtained graph V=G∪{z} is homotopy equivalent to G according to definition 2.2. A point $v_i$ belonging to S is simple in V because the rim $O_V(v_i)$ of $v_i$ in V is the cone z⊕$O_G(v_i)$, i.e. a contractible graph according to theorem 2.1(d). Therefore, every point belonging to S can be deleted from V. The obtained graph H=V-S=(G-S)∪{z} is homotopy equivalent to V, and therefore, to G. This completes the proof. □

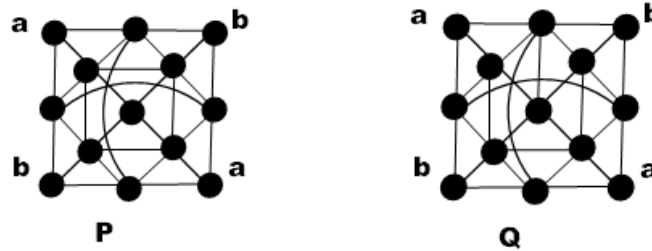

Figure 5. Compressed homotopy equivalent graphs P and Q are not isomorphic.

The deletion of simple points, simple edges and the contraction of simple subgraphs is a natural way to reduce the number of points and edges in a graph. As an example, consider the join W=G⊕H of a contractible graph G and a graph H. H is a simple subgraph of W because B(H)=G⊕H is a contractible graph according to theorem 4.1. Therefore, H can be contracted to a point z, and the obtained graph V=(W-H))∪{z}=G∪{z}=G⊕z is contractible.

**Definition 4.4**
    A graph is called compressed if it contains no simple point, simple edge and simple subgraphs.

The following corollary is a direct consequence of theorem 4.4.

**Corollary 4.1.**
    A graph can be converted to a compressed form by sequential deleting simple points and edges and contracting simple subgraphs.

Compressed graphs which are digital n-manifolds were studied in [18]. Naturally, a graph which has the minimal number of points and edges among all homotopy equivalent graphs is compressed. Fig. 4 shows compressed graphs, which are minimal digital one-, two-, and three-dimensional spheres $S^1$, $S^2$ and $S^3$, a minimal digital two-dimensional torus T, and a minimal digital two-dimensional projective plane P. Notice that homotopy-equivalent compressed graphs are not necessarily isomorphic. As an example, consider graphs P and Q depicted in fig. 5. P and Q are compressed digital 2-dimensional projective planes with equal number of points and edges. They are not isomorphic. Each of them can be considered as a basic graph of the given homotopy class.

**5. Classification of graphs by the complexity, the weight and basic graphs**

Our purpose now is to reduce the order ( number of points) and the size (number of edges) in a graph G by using contractible transformations, which retain the homotopy type of G and to classify graphs by their basic graphs. One can use various tools for the classification of graphs. Usually, objects are classified up to an appropriate equivalence. One path based on the complexity and homotopy equivalence of graphs is considered below. Here it means to define a set of compressed graphs such that any graph is homotopy equivalent to one of them. It is clear, that the classification of graphs in the context of this paper is a computational problem, which requires large amount of memory and computational resources. A contractible graph can be converted into a compressed one-point graph. A graph, which is a digital n-sphere, can be converted to compressed form which is a minimal n-sphere with 2n+2 points [6, 7, 8]. A compressed graph is minimal, i.e. the order and the size of it cannot be reduced by contractible transformations. A compressed graph can be considered as a basic graph of homotopy equivalence class, which this graph belongs to. As mentioned above, homotopy equivalent compressed graphs are not necessarily isomorphic (see graphs P and Q in fig. 5). If there are several homotopy equivalent compressed graphs (with the equal order and size) one of them can be chosen as the basic graph of a given homotopy equivalence class.

It is interesting to stress that each homotopy equivalence type contains an infinite number of graphs. If we find a compressed graph within a given homotopy type, it will simplify many computational tasks in applications where graphs are associated with digital images. Fig. 4 depicts compressed graphs which are digital 1-, 2- and 3-spheres $S^1_1$, $S^2$ and $S^3$, a compressed digital 2-dimensional torus T, and a compressed digital 2-dimensional projective plane P.

Digital n-manifolds have been classified in paper [18]. Using analogy with the results of [18], we give the following definition.

**Definition 5.1**

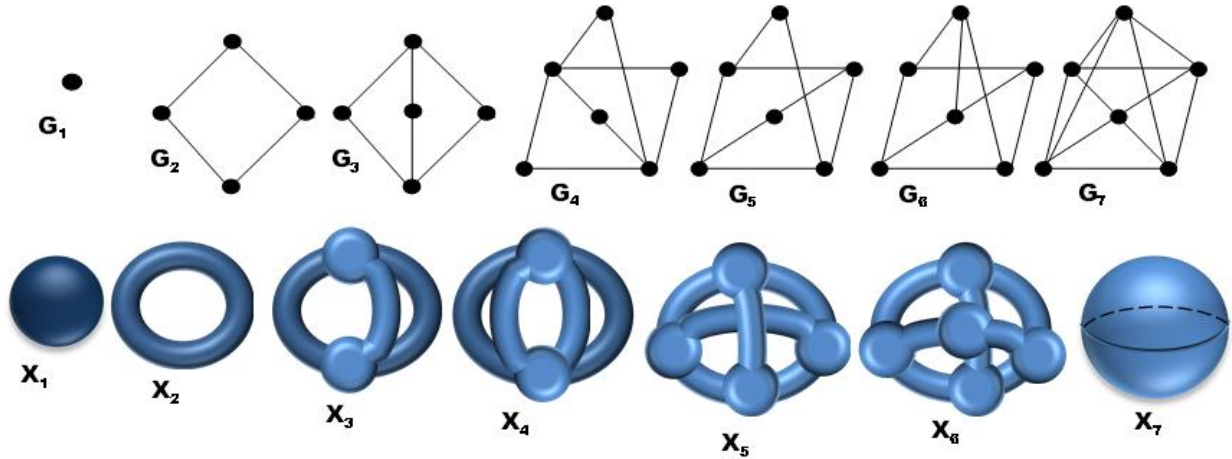

Figure 6. $G_1$-$G_7$ are basic graphs of homotopy equivalence classes with complexity N<7. $G_2$ is a minimal digital 1-sphere, $G_7$ is a minimal digital 2-sphere. Continuous spaces $X_1$-$X_7$ are continuous counterparts of $G_1$-$G_7$.

Let H be a graph and G be a compressed graph homotopy equivalent to H.
- G=bg(H) is called a basic graph for H and the for homotopy equivalence class F containing H and G.
- The complexity of H denoted as com(H) is the order (number of points) of G, com(H)=com(G).
- The weight of H denoted as w(H) is the size (number of edges) of G, w(H)=w(G).

Basic graphs with complexity N<7 are shown in fig. 6. If com(E)≠com(D) or w(E)≠w(D) then graphs E and D are not homotopy equivalent. If com(E)=com(D) and w(E)=w(D) then E and D may be homotopy equivalent or may not be homotopy equivalent. Continuous spaces $X_1$,…$X_7$ are continuous counterparts of basic graphs $G_1$,…$G_7$.

The following algorithm lets us to construct basic graphs of homotopy equivalence classes.
- For a given positive integer N, build a set A of all compressed graphs of order N.
- Separate A into subsets $A_1$,…$A_s$ of homotopy equivalent graphs.
- Pick out from every $A_i$ a graph $G_i$ and call it a basic graph of the homotopy equivalence class. The complexity of this class is equal to N, and the weight is equal to $w(G_i)$.

More detailed classification can be made by using the number of loops and so on. As an example, table 1 contains all basic graphs for complexity N≤6.

| N=1 | N=4 | N=5 | N=6 |
|---|---|---|---|
| $G_1$ | $G_2$ | $G_3$ | $G_4(G_5)$, $G_6$, $G_7$ |

Table 1 Basic graphs of homotopy equivalence classes for complexity≤6.

Graphs $G_4$ and $G_5$ in fig. 6 are homotopy equivalent and any of them can be taken as the basic graph. Similarly, spaces $X_4$ and $X_5$ are homotopy equivalent.

Notice that with no loss of generality in this paper, we study only connected graphs. As one can see from table 1, for N=2,3 there is no basic connected graphs at all. All graphs shown in fig. 6 and graphs $S^1_1$, $S^2$, $S^3$, P and T in fig. 4 are basic graphs of homotopy equivalent classes.

Thus, we can use three elements for the classification of a graph H
- The basic graph G=bg(H) for H.

- The complexity of H com(H)=com(G) (a positive integer).
- The weight of H w(H)=w(G) (a positive integer).

**Examples of applications**

- According to [1], the shape of most macroscopic brain structures can be continuously deformed into a sphere. Then the digital image of this object is a graph, which must be homotopy equivalent to a minimal digital 2-sphere $G_7$ depicted in fig. 6 if the segmentation techniques are accurate and correct. If a continuous object topologically is a torus then its digital image must be homotopy equivalent to a minimal digital torus shown in fig. 4.
- In biology, there is correlations between biocochemical properties of organic molecules ad the topological structure of their molecular graphs (see [19]). A molecular graph can be transformed into its basic graph with the same topological structure as the given molecular graph.

**Conclusion**

The connection is emphasized linking homotopy equivalence of continuous objects and homotopy equivalence of graphs representing these objects. This connection is based on computer experiments. We define a simple curve on a graph and a simply connected graph, and show that a simply connected graph contains a contractible subgraph with the boundary, which is a simple curve.
We show that the join of two graphs is a contractible graph if one of graphs is contractible.
This paper introduces the classification of graphs based on homotopy equivalence and contractible transformations of graphs. Each homotopy equivalence class is characterized by the complexity, the weight and the basic graph. All graphs belonging to a given homotopy equivalence class are homotopy equivalent to each other, have identical topological attributes and are represented by a basic graph with the minimal number of points and edges. Traditionally, the order and the size of the basic graph are called the complexity and the weight of a given homotopy equivalence class and any graph belonging to this class.
A method is presented to compute the complexity, the weight and the basic graph for each homotopy equivalence class. The complexity, the weight and the basic graph can be found by deleting simple points and edges and by contracting simple sets of points in any graph belonging to a given homotopy equivalence class.

**Acknowledgements**

The author would like to thank the anonymous referees for their useful comments and suggestions.